\documentclass[aps,prb,twocolumn,floatfix,superscriptaddress,showpacs]{revtex4}
\pdfoutput=1
\usepackage{graphicx,epsf}
\usepackage{amsmath}
\usepackage{verbatim}
\usepackage{float}
\usepackage{color}

\newcommand{\yrs} {YbRh$_2$Si$_2$}

\newcommand{\TK}    {T_{\rm K}}
\newcommand{\TKO}    {T_{\rm K}^{(1)}}
\newcommand{\Tcoh}  {T_{\rm coh}}

\newcommand{\w}     {\omega}

\newcommand{\bea}{\begin{eqnarray}}
\newcommand{\eea}{\end{eqnarray}}
\newcommand{\beq}{\begin{equation}}
\newcommand{\eeq}{\end{equation}}
\newcommand{\benu}{\begin{enumerate}}
\newcommand{\enu}{\end{enumerate}}

\newcommand{\om}{\omega}

\newcommand{\ep}{\epsilon}

\newcommand{\si}{\sigma}

\newcommand{\ham}{\mathcal{H}}

\newcommand{\cda}{c^{\dagger}}

\newcommand{\fda}{f^\dagger}
\newcommand{\bk}{{\bf k}}

\newcommand{\ek}{\epsilon_{\bk}}
\newcommand{\nn}{\nonumber \\}

\begin{document}

\title{
Doping-induced band shifts and Lifshitz transitions in heavy-fermion metals
}

\author{Adel Benlagra}
\affiliation{Institute for Theoretical Physics, ETH Z\"urich,
8093 Z\"urich, Switzerland}
\affiliation{Institut f\"ur Theoretische Physik, Technische Universit\"at Dresden,
01062 Dresden, Germany}
\author{Matthias Vojta}
\affiliation{Institut f\"ur Theoretische Physik, Technische Universit\"at Dresden,
01062 Dresden, Germany}

\date{\today}

\begin{abstract}
For some heavy-fermion compounds, it has been suggested that a Fermi-surface-changing
Lifshitz transition, which can be driven, e.g., by varying an applied magnetic field, occurs
inside the heavy-fermion regime. Here we discuss, based on microscopic calculations, how
the location of such a transition can be influenced by carrier doping. Due to strong
correlations, a heavy band does not shift rigidly with the chemical potential.
Intriguingly, we find that the actual shift is determined by the {\em interplay} of heavy
and additional light bands crossing the Fermi level: doped carriers tend to populate
heavy and light bands equally, despite the fact that the latter contribute a small
density of states of excitations only.
This invalidates naive estimates of the transition shift based on the low-temperature
specific heat of the heavy Fermi liquid. We discuss applications of our results.
\end{abstract}

\maketitle


\section{Introduction}

Quantum phase transitions in metals are an active area of research.\cite{ssbook,hvl} A
large fraction of the experimentally studied cases occur in heavy-fermion metals, where
strongly localized ($f$) electrons coexist with more itinerant conduction ($c$)
electrons.\cite{hewson,colemanrev} Quite often, the measured deviations from Fermi-liquid
behavior appear incompatible with predictions from standard quantum critical
theories.\cite{si_rev10,stewart01,hvl} At present, it is unclear whether this is related
to yet unexplored types of criticality, to the influence of quenched disorder, or to the
presence of non-universal energy scales preventing to reach the asymptotic critical
regime. Given the smallness of the effective bandwidth of heavy fermions, which is set by
the single-impurity Kondo temperature $\TKO$, it is sometimes difficult to separate
single-particle (i.e. band-structure) effects from those of collective fluctuations.

Recent experiments, utilizing both quantum oscillations and magnetotransport, aim at
detecting Fermi-surface changes across quantum phase transitions, with
\yrs,\cite{paschen04,friede10},
CeRhIn$_5$,\cite{Shishido,Onuki}
CeRh$_{1-x}$Co$_x$In$_5$,\cite{goh}
CeRu$_2$Si$_2$,\cite{daou06,daou12}
and URhGe \cite{yelland} being prominent examples.
Such changes are expected for density-wave transitions where the lattice translation
symmetry gets broken,\cite{ssbook,hvl,norman} for Lifshitz transitions, i.e., transitions
in the topology of the Fermi surface without symmetry breaking,\cite{Lifshitz,Yamaji} and
for the more exotic Kondo-breakdown, or orbital-selective Mott, transitions, where the
heavy quasiparticles themselves cease to
exist.\cite{coleman01,si01,senthil,paul,pepin,osmott_rev,lifkb_note}

A feature of standard Lifshitz transitions is that they can be easily accessed by carrier
doping, which simply shifts the chemical potential, whereas they couple to
pressure only indirectly (via distortions of the quasiparticle dispersion) -- the latter
is in contrast to expectations for both density-wave and Kondo-breakdown transitions.
For \yrs, where a QCP is reached by applying a small magnetic field,\cite{gegenwart}
carrier-doping experiments were proposed in Ref.~\onlinecite{HV11} in order to
discriminate between different transition scenarios:
For bandstructure-related transitions, such as Lifshitz transitions, carrier doping -- as
opposed to isoelectronic doping or pressure -- may be used to directly tune the
transition field. This proposal rests on the idea that the heavy-fermion band responsible
for a potential Lifshitz transition will be shifted relative to the Fermi level upon
doping, while approximately preserving its shape.

This idea prompts two questions, relevant for any Lifshitz transition in heavy-fermion
metals:
(i) How does such a doping-induced band shift occur, given the fact that the
Abrikosov-Suhl resonance (which is responsible for heavy-band formation) is quite
generically pinned to the Fermi level?
(ii) Can one quantitatively estimate the shift, e.g., using the amount of doping and the
low-energy density of states (DOS) of Fermi-liquid excitations (obtained from the
low-temperature specific heat) as input?

In this paper, we shall answer these questions through explicit microscopic calculations
for the Fermi-liquid regime of heavy-fermion metals.
Our findings can be summarized as follows:
(a) Features of heavy-fermion bands, like band edges or van-Hove singularities,
indeed shift upon varying the chemical potential, but they do so much slower than those
of uncorrelated bands. More precisely, their shift is suppressed by a factor which
roughly equals $Z^{-1} = m^\ast/m$ but also depends on further microscopic details. Here,
$m^\ast$ is the effective mass and $Z$ the quasiparticle weight of the heavy carriers.
(b) Since the low-energy DOS of excitations is enhanced by the same factor $Z^{-1}$, an
estimate of the band shift via the specific-heat coefficient, $\gamma=C/T$, would be
appropriate (at least approximately) {\em if} the heavy band under consideration would be
the only band crossing the Fermi level.
However, this is often not the case.
(c) In the presence of both heavy-fermion and weakly correlated bands, doped carriers
enter {\em both} types of bands roughly equally -- despite the small specific-heat
contribution of the weakly correlated bands -- because the latter shift faster upon
varying the chemical potential as compared to the heavy bands, which instead tend to be
pinned to the Fermi level. Then, the band shift {\em cannot} be estimated from the specific heat, but also
depends on the number and properties of the uncorrelated bands.

Together, this implies that the shift of heavy-fermion bands with doping is generically
slower than one would estimate from a free-particle picture, i.e., more
carrier doping is required to affect the location of Lifshitz transitions. This is
of potential relevance for recent experiments on Fe-doped \yrs.\cite{gegen12}

We note that our analysis below is restricted to Lifshitz transitions well inside the
heavy-fermion regime. For magnetic-field-driven transitions, this restriction
conservatively implies that the transition field should be small compared to the Kondo
temperature. Otherwise, the applied field leads to a sizeable polarization of the local
moments. Then, a Lifshitz transition could be accompanied by a field-induced
breakdown of Kondo screening, which would require a separate analysis.
Interestingly, a recent numerical study \cite{bercx12} suggests that even for sizeable
moment polarization accompanying heavy-fermion metamagnetism, the concept of a heavy-band
Lifshitz transition remains well-defined.

The remainder of the paper is organized as follows:
In Sec.~\ref{sec:model} we describe the model Hamiltonian to be employed in our
calculations, together with the relevant approximations.
%
Sec.~\ref{sec:heavy} is devoted to the behavior of a single heavy-fermion band under
variations of the chemical potential.
Sec.~\ref{sec:both} then describes the crucial interplay between light and heavy bands.
A discussion of applications closes the paper.


\section{Microscopic modelling}
\label{sec:model}

In this section, we introduce the model and method for our microscopic approach.

\subsection{Periodic Anderson model}

To describe the heavy-fermion system under the influence of doping, we employ a periodic
Anderson model (PAM) \cite{hewson} of hybridized $c$ and $f$ bands,
\begin{eqnarray}
\ham = && \sum_{\bk \si} \left [ \ek \cda_{\bk \si} c_{\bk \si} + \ep_f \fda_{\bk \si} f_{\bk \si} +
V_\bk ( \cda_{\bk \si} f_{\bk \si} + \mbox{H.c.} )  \right ]\nn
&& +  U \sum_i n^f_{i\downarrow} n^f_{i \uparrow},
\label{PAM}
\end{eqnarray}
in standard notation. In what follows, the hybridization between the $c$ and $f$ fermion
bands will be assumed to be local, $V_\bk=V$, and the local Coulomb repulsion between the
$f$-electrons will be taken to be the largest energy scale, $U\to\infty$.

The heavy Fermi-liquid phase of this model can be captured within a local self-energy
approximation,
\begin{equation}
\Sigma_f(\bk, \om) \rightarrow \Sigma_f(\om).
\label{SE_property}
\end{equation}
Being interested in qualitative low-energy properties at $T=0$, we resort to a standard slave-boson
mean-field treatment.\cite{hewson,coleman84}

\subsection{Slave-boson mean-field approximation}

In the $U \rightarrow \infty$ limit, double occupied states for the $f$ electrons are
projected out. The remaining three states of the $f$ orbital can be represented by
spinless bosons $b_i$ and auxiliary fermions $\tilde f_{i\si}$ with a local constraint:
\begin{equation}
\fda_{i \si} \rightarrow \tilde{f}^\dagger_{i \si} b_i,~~
\sum_{\si}  \tilde{f}^\dagger_{i \si}  \tilde f_{i\si} + b^\dagger_i b_i =1.
\label{constraint}
\end{equation}

In the simplest saddle-point approximation, formally justified in a limit where the spin
symmetry of the original model is extended from SU(2) to SU($N$) with
$N\rightarrow\infty$, the bosonic fields $b_i$ are condensed, and the
constraints are implemented by static Lagrange multipliers $\lambda_i$. Moreover, for a
translational invariant saddle point, both fields are uniform, $b_i\equiv b$ and
$\lambda_i\equiv\lambda$. The original interacting PAM is mapped onto a non-interacting
two-band model with a renormalized $f$ level energy,
$\tilde \ep_f = \ep_f + \lambda$,
and a renormalized hybridization $\tilde{V} = b V$. Diagonalization yields two dispersing
bands,
\begin{equation}
E_{\bk}^\pm=\frac{\ek+\tilde \ep_f \pm \sqrt{(\ek-\tilde \ep_f)^2+4 \tilde{V}^2}}{2},
\label{renormalized_bands}
\end{equation}
describing sharp quasiparticles formed as a mixture of $f$ and $c$ degrees of freedom.
This approximation corresponds to a purely real $f$-electron self energy
\begin{equation}
\Sigma_f(\w) = \lambda b^{-2} + (1-b^{-2}) (\w-\ep_f).
\end{equation}
such that $b$ is also a measure of the mass renormalization and weight of the
quasiparticles, $m/m^\ast = Z = b^2$. A low-temperature Kondo (or coherence) scale may be
defined via $\Tcoh = b^2 W$ where $W$ is the half-bandwidth of the conduction band.

The parameters $\lambda$ and $b$ are obtained from the mean-field equations
\begin{subequations}
\label{eq:MF_ALM}
\begin{eqnarray}
V\sum_{{\bf k}\sigma} \langle \bar f_{{\bf k} \sigma}^{\dagger} c_{{\bf
k} \sigma}^{\phantom{\dagger}}+ h.c. \rangle &=& - 2 \mathcal {N} \lambda b , \\
\sum_{{\bf k} \sigma}  \langle \bar f_{{\bf k} \sigma}^{\dagger} \bar f_{{\bf k}
\sigma}^{\phantom{\dagger}}\rangle &=& \mathcal{N} (1-b^2)
\end{eqnarray}
\end{subequations}
where $\mathcal {N}$ is the number of lattice sites.

Detailed numerical studies using dynamical mean-field theory
\cite{grenzebach06,assaad_DMFT} and its cluster generalizations \cite{assaad_CDMFT} have
verified that the effective two-band picture emerging from the slave-boson approximation
qualitatively captures the low-energy physics of the heavy Fermi-liquid phase of the PAM.


\subsection{Band structure}
\label{sec:scen}
\label{sec:narrowmodel}

\begin{figure}[!t]
\centering
\includegraphics[width=8.6cm]{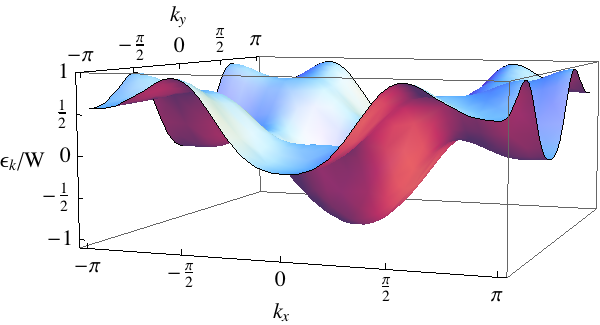}
\caption{The bare band dispersion $\epsilon_\bk$ as defined in
Eq.~\eqref{bare_dispersion}  in the first Brillouin zone. Notice the weakly dispersing
piece near $\bk=(\pi,0)$ and $(0,\pi)$.
}
\label{fig:3Ddispersion}
\end{figure}

As we are interested in Lifshitz transitions well inside the heavy-fermion regime, we
study a situation with a shallow Fermi pocket within the heavy-fermion band. We
generate this simply by ``engineering'' the bare conduction-electron band $\ek$ such that
it has a weakly dispersing portion with local minima and maxima in the middle of the band.
Specifically, we employ a dispersion on a 2d square lattice which includes longer-range hopping
terms:
\begin{eqnarray}
\ek&=&-2t_{10}(\cos k_x + \cos k_y)-2t_{20}(\cos 2k_x+ \cos 2k_y)  \nonumber\\
&-& 2t_{40}(\cos 4k_x+\cos 4k_y) - 4t_{11}\cos kx \cos ky \nonumber\\
&-& 4t_{22}\cos 2k_x \cos 2k_y - 4t_{33}\cos 3k_x \cos 3k_y - t_{00}, \nn
\label{bare_dispersion}
\end{eqnarray}
%
%
with parameters $t_{10}=2 t_{20} = 2.04 \times 10^{-1}, t_{40}=-2.92\times 10^{-2},
t_{11}= 1.02 \times 10^{-2}, t_{22}= 3.40\times 10^{-2}, t_{33}= 6.80 \times
10^{-3}$ and $t_{00}= 2.45 \times 10^{-1}$, which has a weakly dispersing piece near
$\bk=(\pi,0)$ and $(0,\pi)$, as can be seen in Fig.~\ref{fig:3Ddispersion}.
The dispersion's half-bandwidth is $W=1$ which we choose as our energy unit.

In the low-temperature heavy-fermion regime, the shape of the bare $c$ band
\eqref{bare_dispersion} will generate portions of heavy bands which disperse weakly on
the scale $\Tcoh$. Those can be brought near the Fermi level by choosing an
appropriate value of the chemical potential $\mu$.

\begin{figure}[!t]
\centering
\includegraphics[width=8.6cm]{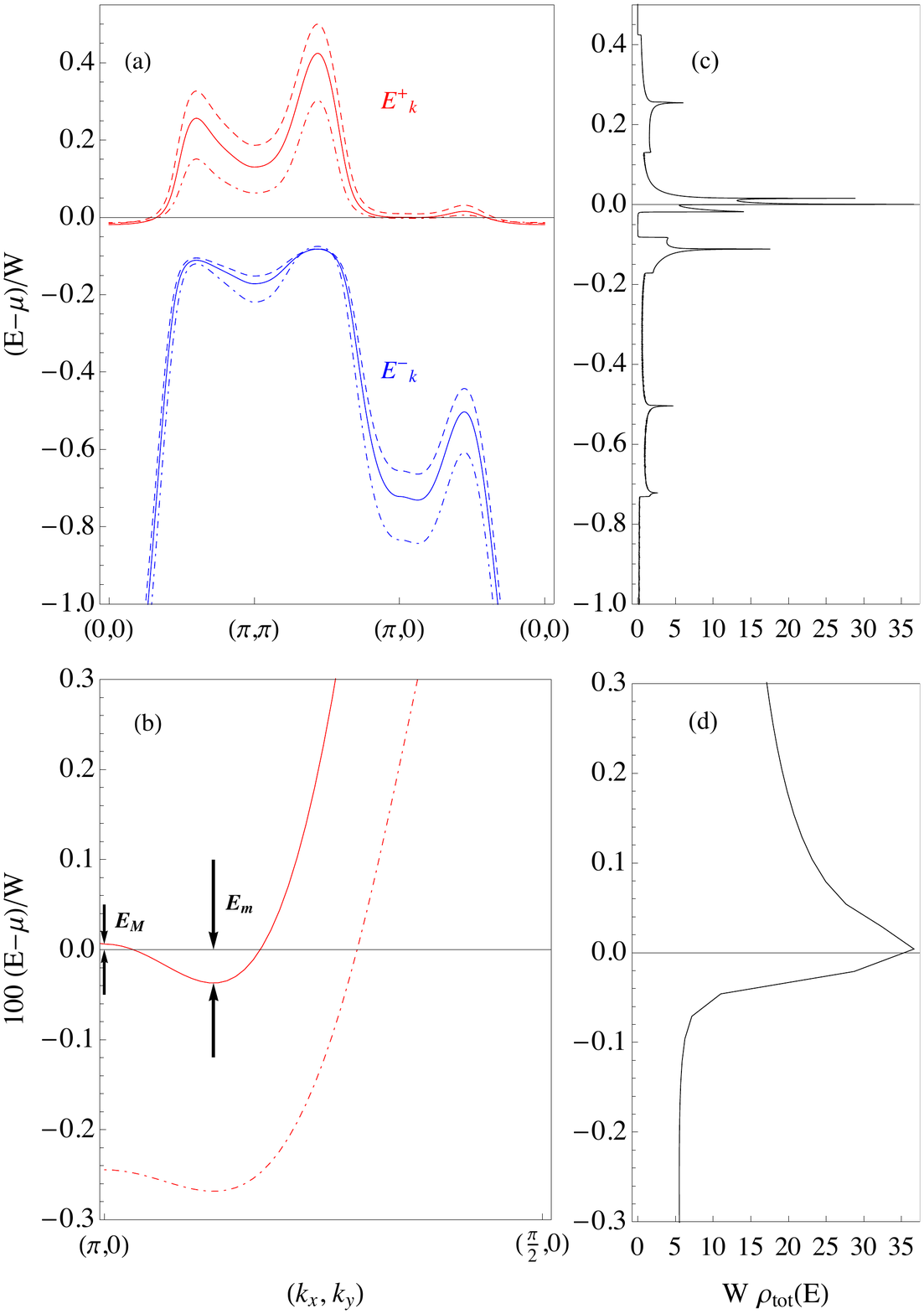}
\caption{
(a,b): Zero-temperature heavy-fermion band structure with a shallow pocket, as obtained
within the slave-boson mean-field approach. (a) shows the full dispersion of the two
hybridized bands along a path in the 2d Brillouin zone for different values of the
chemical potential $\mu$, (b) a zoom near the Fermi level. The depth of the shallow Fermi
pocket is $E_m\equiv E^+_{\bk=(2.75,0)}$ whose doping evolution will be discussed in the
paper. The height of the pocket defines a second scale $E_M\equiv E^+_{\bk=(\pi, 0)}$;
the vHs at $(\pi, 0)$, $(0,\pi)$ produces a peak in the DOS.
(c,d): Corresponding total DOS (per spin) of quasiparticles, $\rho_{\rm tot} =
\rho_c + \rho_{\tilde f}$.
The bare band dispersion is as specified in Eq.~\eqref{bare_dispersion}, furthermore
$\ep_f/W = -9.65$, $\mu/W=0.575$ (resp. $\mu/W=0.499$ for the dashed dispersion and
$\mu/W=0.651$ for the dot-dashed one) and $V/W=1.5$, which results in a filling of
$n_{\rm tot} =  2.33$ (resp. $n_{\rm tot} = 2.23$ and $n_{\rm tot} =  2.45$) and a
quasiparticle renormalization of $1/Z\simeq 99$ (resp. $1/Z\simeq86 $ and $1/Z\simeq
120$). Momentum integrals were performed with $1600^2$ $\bf k$ points, and a Lorentzian
broadening of $10^{-5}W$ was employed for the DOS -- this smears the logarithmic
divergence of the DOS at the vHs.
}
\label{fig:bands}
\end{figure}


An example, with a heavy-fermion band structure derived from the slave-boson
approximation, is shown in Fig.~\ref{fig:bands}a. The weakly dispersing piece,
Fig.~\ref{fig:bands}b, induces a pronounced peak  in the DOS at the Fermi level, on top
of the usual Kondo peak originating from the Abrikosov-Suhl resonance.
(Recall that the width of the latter is roughly given by $WZ$, here $10^{-2}$.)
In fact, the maximum DOS arises from the van-Hove singularity (vHs) at $\bk=(\pi,0)$,
$(0,\pi)$ in the bare dispersion \eqref{bare_dispersion}; this vHs is inherited by both
the upper and lower renormalized bands. Together with the dispersion minimum at
$\bk=(2.75,0)$ this defines a small and shallow Fermi pocket.
Its depth and height are measured by $E_m$ and $E_M$, respectively, see Fig.~\ref{fig:bands}b.

Moving the chemical potential away from the position corresponding to Fig.~\ref{fig:bands}b is
expected to lead to Lifshitz transitions within the heavy-fermion regime;
in particular, $E_m(\mu)=0$ corresponds to the point where the shallow pocket disappears.
Alternatively, this transition is expected to be driven for a single spin species at finite
$E_m$ by applying a Zeeman field.



\section{Carrier doping of heavy-fermion bands}
\label{sec:heavy}

Our main objective is to understand the behavior of heavy-fermion bands upon carrier
doping. We shall therefore study the PAM of Sec.~\ref{sec:model} under variation of the
chemical potential. Specifically, we are interested in a regime where a narrow-band
feature is located near the Fermi level, and we would like to understand how a shallow
Fermi pocket, as in Fig.~\ref{fig:bands}b, evolves upon doping. For this purpose, we will
monitor the pocket depth $E_m$ upon doping.

We note that a variation of the chemical potential will in general have two different
effects: It will cause a change in the occupation numbers and thus a band shift, but it
will also cause a change in the Kondo temperature $\TK$. The latter effect is primarily
determined by the energy dependence of the {\em bare} $c$-electron density of states at
the Fermi level and therefore disconnected from the physics of the shallow bands in the
{\em renormalized} band structure. However, a consistent calculation requires to account
for both effects, because a change in $\TK$ comes with a change in $b$ and, via
Eq.~\eqref{constraint}, a change in the $f$ electron occupation.


\subsection{Quantifying band shifts}

In order to quantify the band-shift phenomenology,
let us consider uncorrelated electrons as a reference. Here, a change in the chemical
potential, $\Delta\mu$, produces a rigid band shift, such that $\Delta E_m = \Delta\mu$.
Moreover, this band shift leads to a change in the occupation number $n$ according to $\Delta n
= 2 \rho(0) \Delta\mu$ at zero temperature where $\rho(0)$ is the DOS (per spin) at the Fermi level.
(Note that $\Delta n/\Delta\mu$ is the electronic compressibility.)
For heavy Fermi liquids, this suggests to focus on the quantities
\begin{equation}
P = \frac{\Delta E_m}{\Delta \mu},~~
Q = \frac{\Delta n_{\rm tot}}{2 \rho_{\rm tot} \Delta \mu},
\label{defPQ}
\end{equation}
where $n_{\rm tot}$ is the number of electrons in both the $c$ and $f$ bands of the model
\eqref{PAM}, and $\rho_{\rm tot}$ denotes the (renormalized) total DOS per spin,
which is related to the $T\to 0$ specific-heat coefficient $\gamma$ by the Fermi-liquid
relation $\gamma=(2\pi^2/3)\rho_{\rm tot}$. 
Thus the ratio
\begin{equation}
\frac{P}{Q} = \frac{3}{\pi^2} \frac{\Delta E_m \gamma}{\Delta n_{\rm tot}}
\label{poq}
\end{equation}
relates the experimentally accessible observables $\Delta n_{\rm tot}$ and $\gamma$ to
the band shift $\Delta E_m$, which itself would, e.g., correspond to a shift in the
transition field of a Zeeman-induced Lifshitz transition associated with the Fermi
pocket.

While both $P$ and $Q$ equal unity for uncorrelated bands, we will show below that both
quantities are suppressed by roughly $Z=m/m^\ast$ in the heavy-fermion regime.
In the effective two-band description, the central effect which causes a deviation from
the rigid-band picture is that the $f$ occupation remains (approximately) fixed (as a
result of strong correlations) upon varying the chemical potential.

For the evaluation of $Q$ we note that, in the slave-boson approximation, the
renormalized total DOS, determining the specific-heat coefficient, is simply given by
the total quasiparticle DOS, $\rho_{\rm tot} = \rho_c + \rho_{\tilde f}$.

\begin{figure}[!t]
\centering
\hspace{-0.2cm}\includegraphics[width = 8.cm]{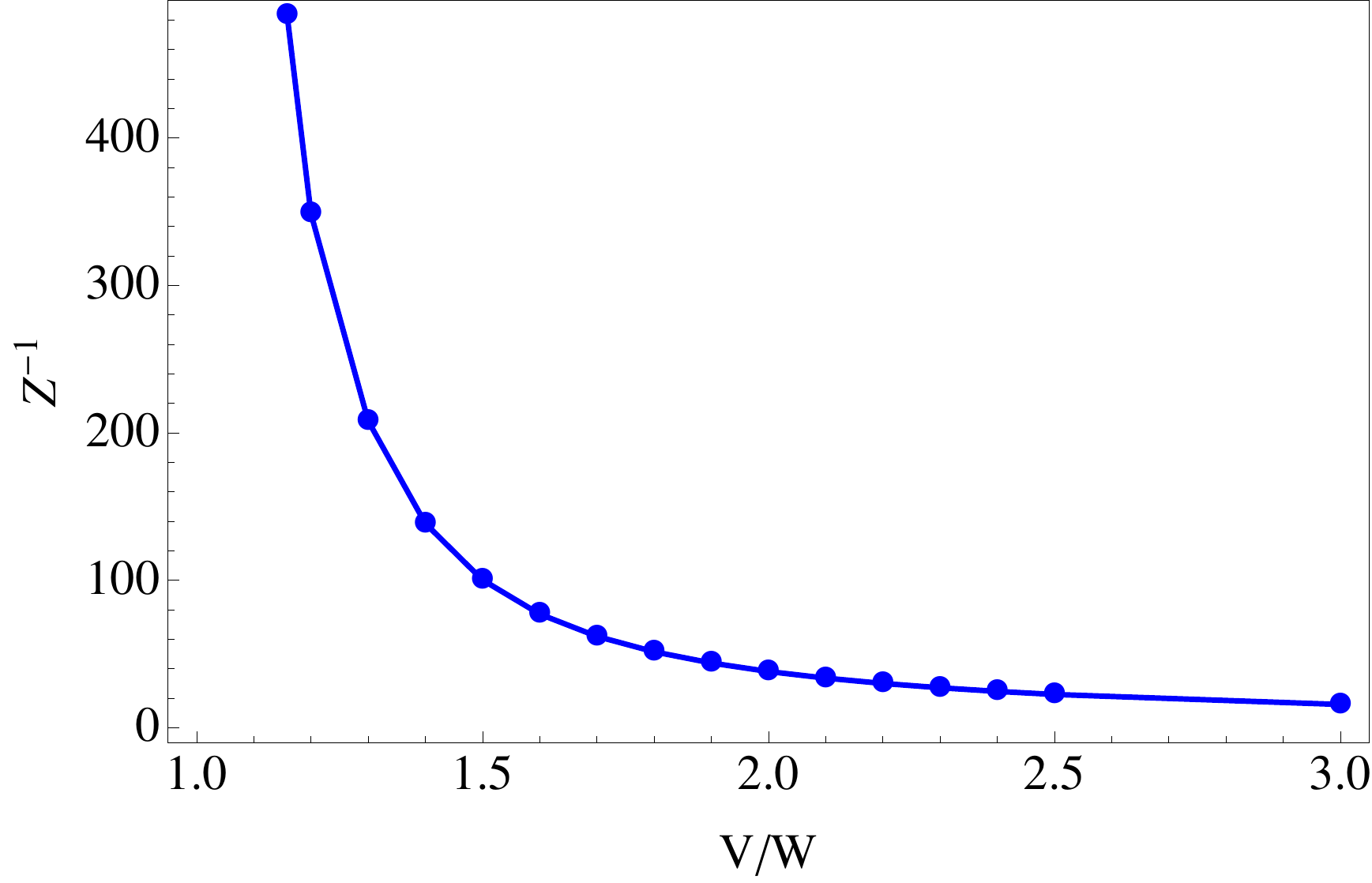}
\vspace{-0.2cm}
\caption{
Evolution of the mass renormalization $Z^{-1}=m^\ast/m$ with the hybridization $V$, as
obtained from the slave-boson approximation. Here, $\ep_f/W = -9.65$ is kept fixed and
$\mu$ is adjusted to $\mu_0(V)$ such that $|E_m|=|E_M|$, see text for details. $Z(V)$ is
dominated by the exponential dependence\cite{hewson} of the single-impurity Kondo
temperature $\TKO$ on $V$.
}
\label{fig:zvsv}
\end{figure}

\subsection{Slave-boson results}

We now turn to our numerical results from the slave-boson
mean-field approximation, using the bare $c$ band structure as
described in Sec.~\ref{sec:narrowmodel}. A variation of the Kondo temperature (or mass
renormalization) is achieved by fixing the bare $f$ level energy $\ep_f$ while varying
the hybridization strength $V$.
For each parameter set, the chemical potential $\mu$ is adjusted to a value $\mu_0(V)$
such that the Fermi pocket around $\bk=(2.75,0)$ is present in the upper band $E_\bk^+$,
with $|E_m|=|E_M|$. This prescription leads to band fillings $n_{\rm tot}>2$ which vary
with $V$, but guarantees that the narrow band feature of interest is centered at the
Fermi level. (Alternative choices for $\mu_0(V)$ do not alter the qualitative results.)

Fig.~\ref{fig:zvsv} shows the evolution of $1/Z = m^\ast/m$ with $V$ where $\mu=\mu_0(V)$ for
all $V$. As expected, $1/Z$ strongly increases upon decreasing $V$ -- this
mainly reflects the exponential dependence of the Kondo temperature on $V$: $\ln\TKO
\propto V^2/(\ep_f W)$.

\begin{figure}[!t]
\centering
\vspace*{-3mm}
\includegraphics[width = 8.7cm]{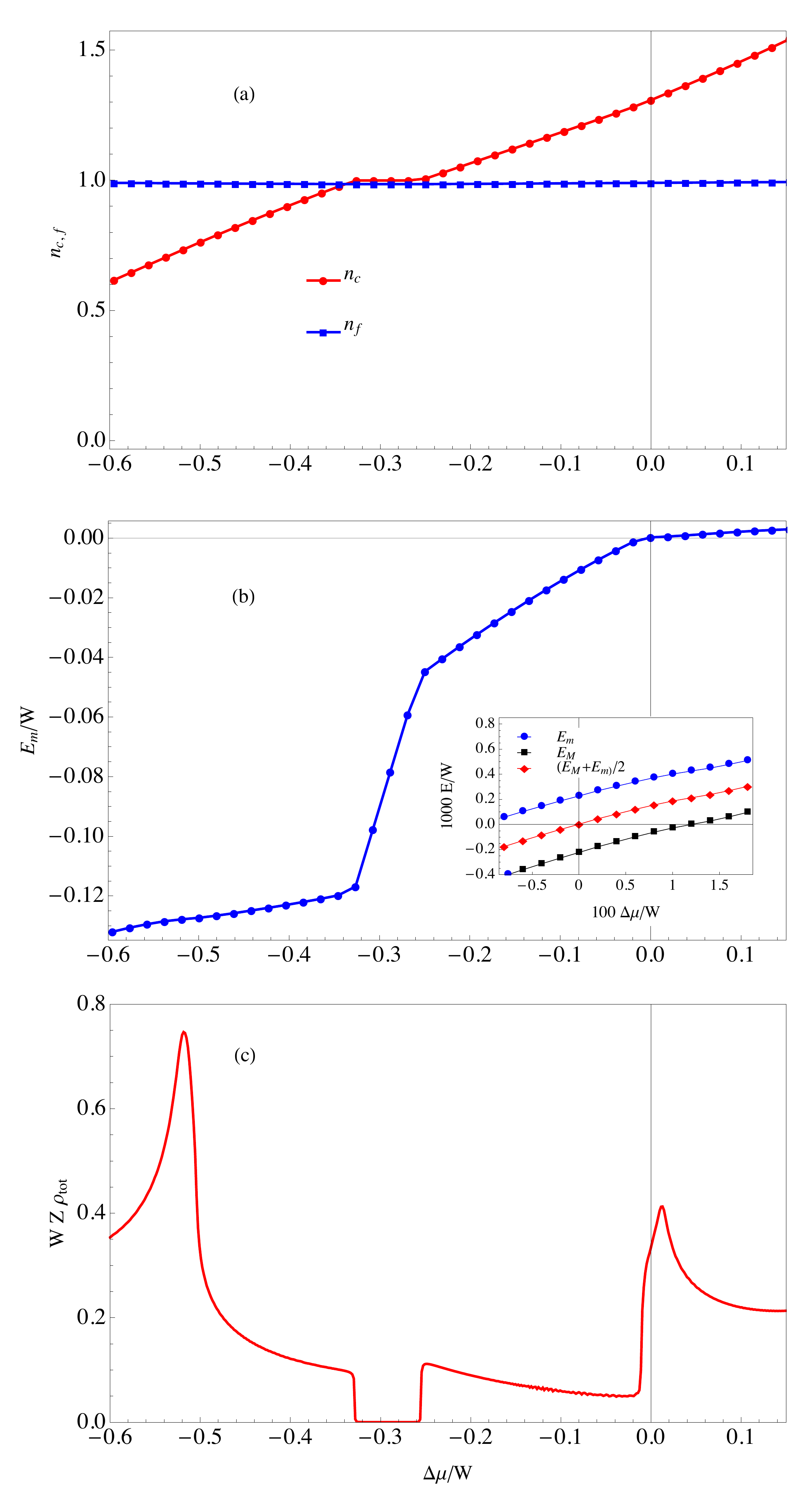}
\caption{
Slave-boson results for
(a) the band fillings $\langle n_c\rangle$ and $\langle n_f\rangle$,
(b) the energy scale $E_m$ measuring the pocket depth, and
(c) the total quasiparticle DOS at the Fermi level normalized by the quasiparticle weight, $\rho_{\rm tot}(0)Z$,
as function of the chemical potential, defined via
$\mu = \mu_0(V) + \Delta\mu$, where $|E_m|=|E_M|$ for $\Delta\mu=0$.
Here, $\ep_f/W = -9.65$, $V/W=1.5$ are kept fixed, and $\mu_0(V)/W\simeq0.569$.
The interval $-0.33<\Delta\mu/W<-0.26$ where $\langle n_c\rangle$, $\langle n_f\rangle$ are
constant corresponds to the Kondo insulator.
Comparing panel (c) with the energy-dependent DOS at {\em fixed} $\mu=\mu_0(V)$ in
Fig.~\ref{fig:bands}d illustrates again that a rigid-band picture is by no means
appropriate.
}
\label{fig:fill}
\end{figure}

Fig.~\ref{fig:fill}a illustrates the variation of the band filling in the vicinity of $\mu=\mu_0(V)$
for fixed $V/W=1.5$.
%
\begin{figure}[!h]
\centering
\vspace*{-0mm}
\includegraphics[width = 7.8cm]{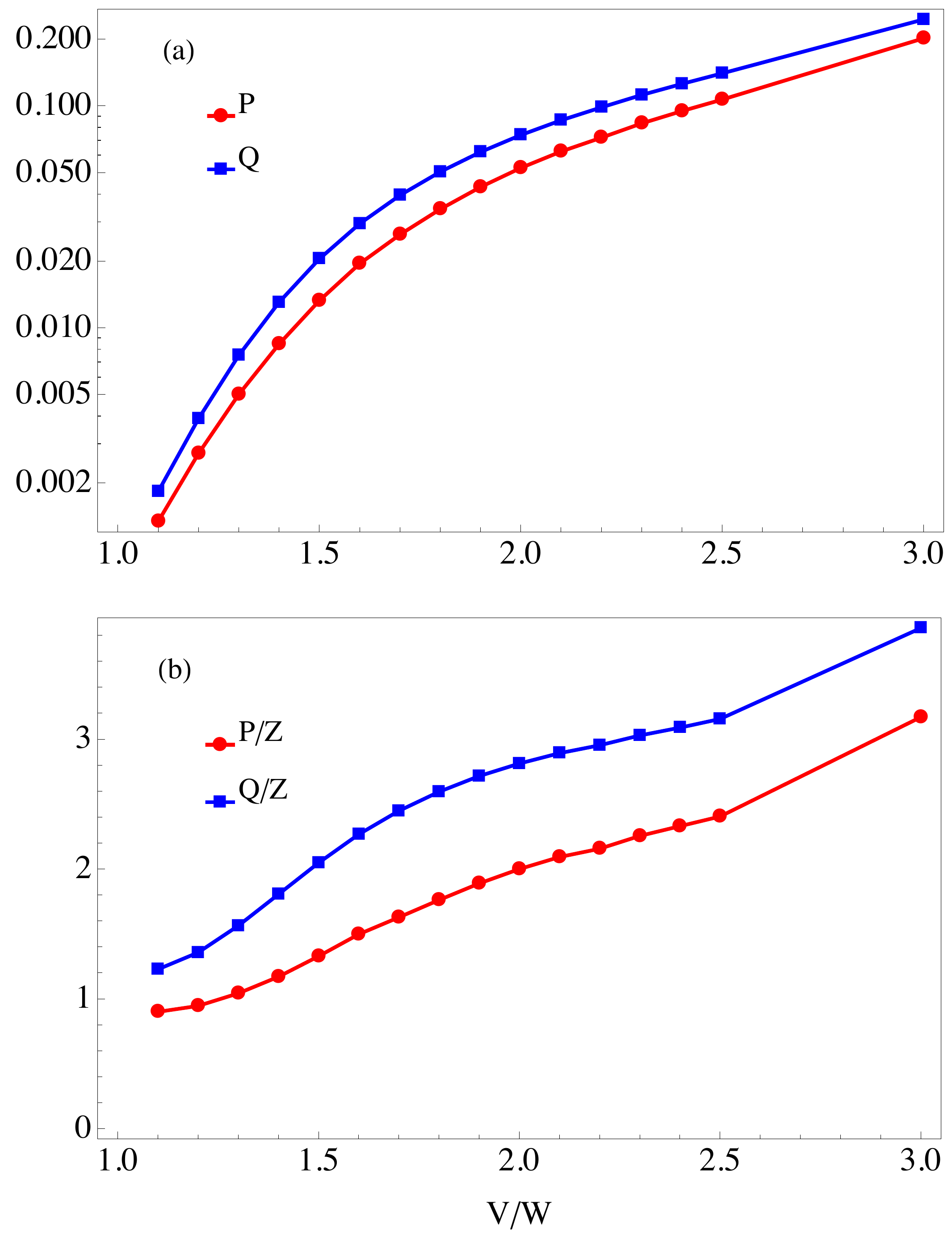}
\caption{
(a) Dimensionless parameters $P$ and $Q$, defined in Eq.~\eqref{defPQ}, as function of
the hybridization $V$, with $\ep_f/W=-9.65 $ and $\mu=\mu_0(V)$ as before.
$P$ and $Q$ characterize the changes of $E_m$ and $\langle n\rangle$ with $\mu$,
respectively, and appear strongly suppressed w.r.t. the non-interacting reference value
1.
(b) Re-scaled parameters $P/Z$ and $Q/Z$, which are now of order unity.
}
\label{fig:PQ}
\end{figure}
In Fig.~\ref{fig:fill}b we show $E_m$, the position of the bottom of the Fermi pocket. It
is clear that this does not follow the imposed chemical-potential variation, but changes
more slowly, i.e., $P\ll 1$. As can be seen in the inset, $E_m$ and $E_M$ have roughly
the same variation in the vicinity of $\mu_0(V)$, such that the width of the shallow Fermi
pocket is approximately preserved for this range of chemical potential.
Fig.~\ref{fig:fill}c also shows the evolution of the renormalized total DOS at the Fermi level. Most
importantly, this DOS is large in units of $(1/W)$, such that it becomes of order unity
upon multiplying by the quasiparticle residue, $W Z \rho_{\rm tot}(0) \sim 1$. Considering
the variation of the filling in Fig.~\ref{fig:fill}a where $\Delta n \sim \Delta\mu/W$,
it is clear that $Q/Z \sim 1$, i.e., $Q \ll 1$.
Results similar to that in Fig.~\ref{fig:fill} can be found for any $V$.

In Fig.~\ref{fig:PQ}, we have collected data for $P$ and $Q$ obtained upon varying $V$.
Both $P$ and $Q$ have been evaluated utilizing small variations of $\mu$ in the vicinity
of $\mu=\mu_0(V)$.

\subsection{Summary}

For the simple two-band heavy fermion system, we have investigated the quantities $P$ and
$Q$, measuring the shift of heavy-band features and the change in carrier density
normalized to the DOS of excitations, respectively, upon changing the chemical potential.
Both $P$ and $Q$, which equal unity for uncorrelated bands, are renormalized downwards by
approximately a factor of $Z=m/m^\ast$ -- this is a natural consequence of the ``pinning''
of the Kondo resonance (and with it the heavy band) to the Fermi level.

Interestingly, the fact that $P$ and $Q$ are renormalized in parallel implies that the
ratio $P/Q$ \eqref{poq} is of order unity, Fig.~\ref{fig:PQ}b.
However, as we show in the next section, this state of affairs changes once the
complexity of real heavy-fermion band structures is accounted for.

\section{Interplay of light and heavy bands}
\label{sec:both}

If a heavy-fermion system displays, in addition to the heavy-electron band discussed so
far, other weakly correlated (i.e. light) bands crossing the Fermi level, then the quantity $Q$ needs
to be re-defined:
\begin{equation}
Q = \frac{\Delta n_{\rm tot,h} + \Delta n_{\rm l}}{2(\rho_{\rm tot,h} + \rho_{\rm l}) \Delta \mu}
\label{fullQ1}
\end{equation}
where the indices $\rm l$ and $\rm h$ refer to the contributions from the light and heavy
bands, respectively (where the ``heavy'' piece includes the two renormalized bands of the
PAM as described in Sec.~\ref{sec:narrowmodel}, with one of them crossing the Fermi
level). Provided that $\rho_{\rm tot,h} \gg \rho_{\rm l}$ and using $\Delta n_{\rm l} = 2
\rho_{\rm l} \Delta\mu$ the above equation reduces to
\begin{equation}
Q =
\frac{\Delta n_{\rm tot,h}}{2\rho_{\rm tot,h} \Delta \mu} + \frac{\rho_{\rm l}}{\rho_{\rm tot,h}}
= Q_{\rm h} + \frac{\rho_{\rm l}}{\rho_{\rm tot,h}}
\label{fullQ2}
\end{equation}
where $Q_{\rm h} \sim Z$ is the heavy-band $Q$ calculated in Sec.~\ref{sec:heavy}.
Assuming the bare DOS in a light band to be comparable to that of the $c$ band forming
the heavy fermions, we have $\rho_{\rm l}/\rho_{\rm tot,h} \sim Z$, such that both
contributions in Eq.~\eqref{fullQ2} are of the same order of magnitude. Moreover,
$\rho_{\rm l}$ scales with the number of light bands crossing the Fermi level.

\begin{figure}[!t]
\centering
\includegraphics[width = 8cm]{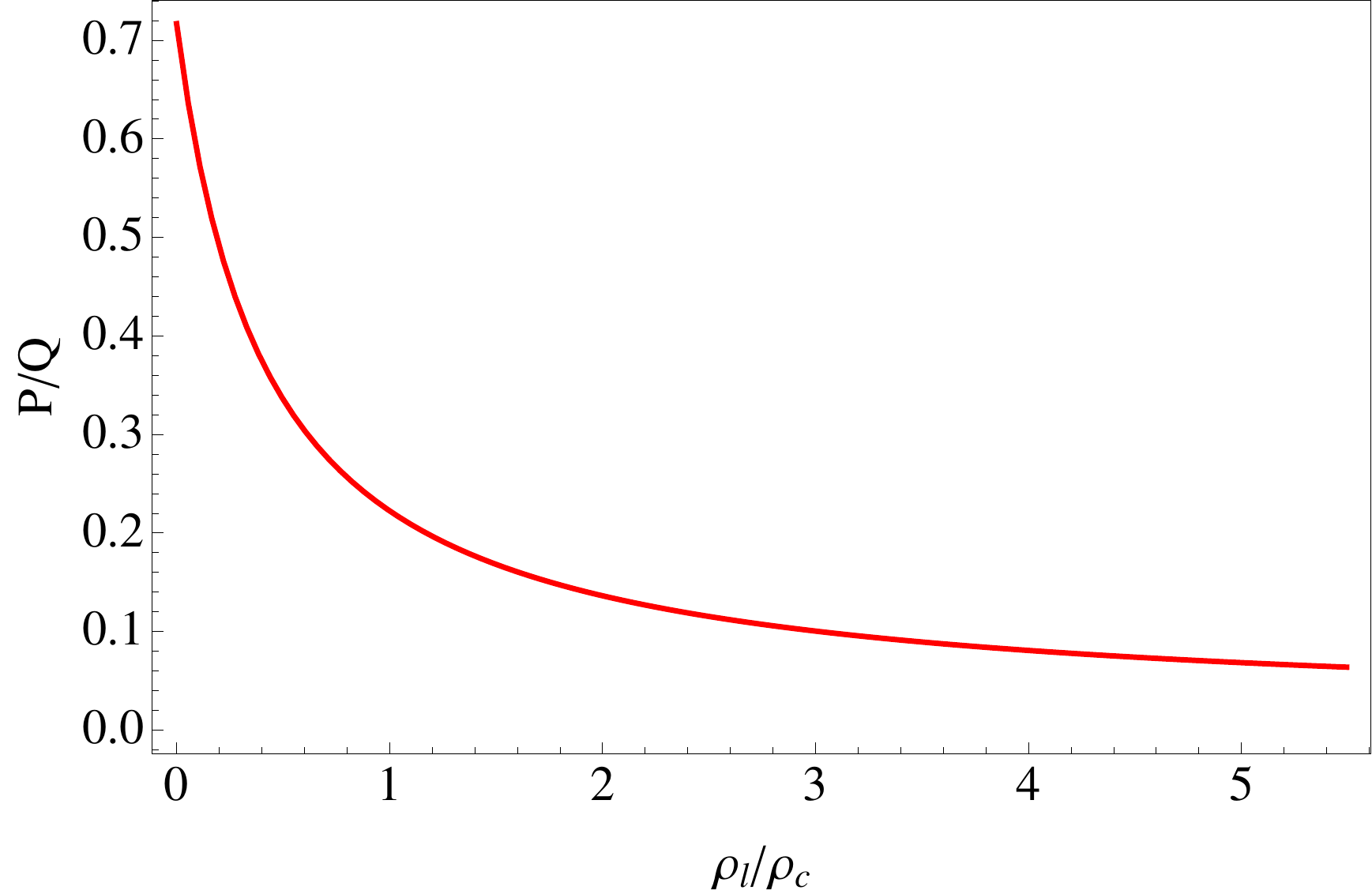}
\caption{
Evolution of the ratio $P/Q$, defined in Eq.~\eqref{poq}, with the addition of an
increasing number of light bands according to Eq.~\eqref{fullQ1}. $\rho_l$ scales with
the number of light bands crossing the Fermi level. The heavy-band parameters are the
same as in Fig.~\ref{fig:PQ} with $V/W=1.5$, and $\rho_c$ refers to the $c$-electron part of the DOS per spin $\rho_{tot,h}$
of the two-band system described by the PAM with the same heavy-fermion band parameters.
}
\label{fig:light_bands}
\end{figure}

This implies that the contribution to $Q$, i.e., to the number of doped carriers, from the
light bands is {\em not} negligible, despite the light bands' DOS being much smaller than
that of the heavy band.
Inserting numbers, we find that $P/Q$ can easily reach values down to 0.1--0.2, see
Fig.~\ref{fig:light_bands}, with small values occurring if the weakly correlated bands
happen to have a larger DOS than the bare heavy-fermion $c$ band. Thus, for
a given number of doped carriers, the heavy-band features such as band edges and vHs
shift much slower than estimated via the specific heat, the main reason being that the
carriers enter both heavy and light bands simultaneously, because the light uncorrelated
bands are more susceptible to changes in the chemical potential.

\section{Conclusions}

We have investigated the effect of carrier doping on quasiparticle bands of heavy
Fermi liquids, with the goal of quantifying the shift of band-structure features such as band
edges and van-Hove singularities (of shallow pockets) with doping.

We have found, not unexpectedly, that a rigid shift of heavy-fermion bands does not
occur, due to the pinning of the Abrikosov-Suhl resonance to the Fermi level. This
reduced band shift is paralleled by a reduced compressibility: The charge doped into a
two-band heavy Fermi liquid, as described by the standard periodic Anderson model, is
much smaller than its large DOS of Fermi-liquid excitations would suggest. Interestingly, these
two effects tend to cancel when it comes to estimating the band shift from the doped
charge via the specific-heat coefficient, Eq.~\eqref{poq}.

For real materials another issue comes into play: Because of the reduced compressibility
of the heavy bands, additional weakly correlated (i.e. light) bands cannot be neglected
for the doping process, despite their small contribution to the specific heat. As a
result, doped carriers populate both heavy and light bands, such that knowledge of the
specific heat is not sufficient to estimate the band shift.

We now briefly discuss the application of our results to \yrs. Here, recent
carrier-doping experiments using Fe substituting for Rh indicate that the transition
field $B^\ast$, as marked by the termination of the so-called $T^\ast$ line at $T=0$,
(Refs.~\onlinecite{paschen04,gegenwart07,friede10}) can be tuned by doping.
On the one hand, one can estimate the shift of band-structure features in a free-electron model where
$P=Q=1$. Using an approximate $\gamma=2$\,J/mol\,K$^2$ yields a shift of $\Delta E_m = 0.06$\,meV for 5\%
Fe doping.
On the other hand, $B^\ast$ has moved by 30\,mT for this doping.\cite{gegen12} Using the
high-field $g$ factor of 3.6 this converts into $\Delta E_m = 0.006$\,meV. Therefore, if
the transition is driven by band-structure features, consistency requires $P/Q\approx
0.1$. While this small value could be due to an unexpectedly large DOS of the weakly
correlated bands, it is more likely that non-local correlation effects not captured in
the present calculation become relevant. One candidate is a low-temperature enhancement
of the $g$ factor due to incipient ferromagnetism \cite{gegen05} -- this would then imply larger $\Delta
E_m$ and hence larger $P/Q$.
Further experiments using different dopants may shed light onto this issue: For a
transition driven by band-structure effects, one would expect a shift of $B^\ast$ which,
to leading order, depends on the number of doped carriers only, i.e., Fe and Ru doping should
have a similar effect on $B^\ast$.


\acknowledgments

The authors acknowledge fruitful discussions with F. Assaad, M. Brando, S. Friedemann, P.
Gegenwart, and Y. Tokiwa.
This research has been supported by the DFG through GRK 1621 and FOR 960.


\end{document}